# A Generalized Information Formula as the Bridge between Shannon and Popper

Chenguang Lu[1]

Independent Researcher, survival99@hotmail.com

**Abstract.** A generalized information formula related to logical probability and fuzzy set is deduced from the classical information formula. The new information measure accords with to Popper's criterion for knowledge evolution very much. In comparison with square error criterion, the information criterion does not only reflect error of a proposition, but also reflect the particularity of the events described by the proposition. It gives a proposition with less logical probability higher evaluation. The paper introduces how to select a prediction or sentence from many for forecasts and language translations according to the generalized information criterion. It also introduces the rate fidelity theory, which comes from the improvement of the rate distortion theory in the classical information theory by replacing distortion (i.e. average error) criterion with the generalized mutual information criterion, for data compression and communication efficiency. Some interesting conclusions are obtained from the rate-fidelity function in relation to image communication. It also discusses how to improve Popper's theory.

## 1 Introduction

Although Shannon's information theory is successful for electrical communication, it does not deal with semantic information [1]. Semantic information measures have been discussed for long time [2], [3], [4], [5]. However, no formula can be properly used to measure the information of a prediction like "Tomorrow will be rainy" or "Temperature is about 10 °C".

Twenty years ago, I set up a symmetrical model of color vision with four pairs of opponent colors instead of three pairs as in the popular zone model [6]. To prove that the more unique colors we perceive, the higher discrimination to lights our eyes have, and hence the more information we can obtain, I researched information theory. Similarly, information conveyed by color vision is also related to semantics or meaning of symbols. From 1989 to 1993, I found the generalized information formula [7], [8] and published the monograph [9] on generalized information. Later, I wrote some papers for further discussion [10], [11] and published a monograph on portfolio and information value [12].

This paper will focus on the generalized information criterion for selecting one from several sentences, or predictions, and on the rate fidelity theory, which is an improved version of classical rate distortion theory, for data compression and communication efficiency.

## 2 Deducing Generalized Information Formula

First we consider the information provided by predictions such as "The growth rate of GDP of this year will be 8%".

Let $X$ denote the random variable taking values from set $A=\{x_1, x_2...\}$ of events, such as the growth rates or temperatures, $Y$ denote the random variable taking values from set $B=\{y_1, y_2...\}$ of sentences or predictions. For each $y_j$, there is a subset or fuzzy subset $A_j$ of $A$ and $y_j = "x_i \in A_j"$.

In the classical information theory, information provided by $y_j$ about $x_i$ is

$$I(x_i; y_j) = \log \frac{P(x_i | y_j)}{P(x_i)}. \qquad (1)$$

---

[1] see his home page: http://survivor99.com/lcg/english for more articles.

Yet, in linguistic communication we only know meaning of a sentence or a prediction instead of the condition probability $P(x_i|y_j)$. Fortunately, we can deduce the condition probability $Q(x_i/A_j)$ of $x_i$ while condition $x_i \in A_j$, which means that $y_j =$ "$x_i \in A_j$" is true, by Bayesian formula

$$Q(x_i | A_j) = \frac{Q(A_j | x_i)P(x_i)}{Q(A_j)}, \quad (2)$$

where

$$Q(A_j) = \sum_i P(x_i)Q(A_j | x_i). \quad (3)$$

Replacing $P(x_i|y_j)$ with $Q(x_i|A_j)$ in (1), we have the generalized information formula:

$$I(x_i; y_j) = \log \frac{Q(x_i | A_j)}{P(x_i)} = \log \frac{Q(A_j | x_i)}{Q(A_j)}, \quad (4)$$

which is illustrated by Fig. 1.

Note that the most important thing is that generally $Q(x_i|A_j) \neq P(x_i |y_j)$ because $P(x_i|y_j)= P(x_i|$"$x_i \in A_j$"$)=P(x_i |$"$x_i \in A_j$" is reported); yet, $Q(x_i|A_j)= P(x_i|x_i \in A_j)=P(x_i |$"$x_i \in A_j$" is true). The $y_j$ may be an incorrect prediction or a lie; yet, $x_i \in A_j$ means that $y_j$ must be correct. If they are always equal, then generalized information formula (4) will become the classical information formula (1).

The generalized information formula can measure not only semantic information, but also sensory information. Let $X$ denote one of monochromatic lights, $Y$ denote the corresponding color perception, $A_j$ denote fuzzy set, which includes all $x_i$ that are confused with $x_j$, of $A$, and $Q(A_j|x_i)$ denote the confusion probability of $x_i$ with $x_j$. Then, a color perception can be regarded as a sentence $y_j =$ "The color $x_i$ is about $x_j$". Hence, the generalized information formula can also be used to measure the information of a color perception.

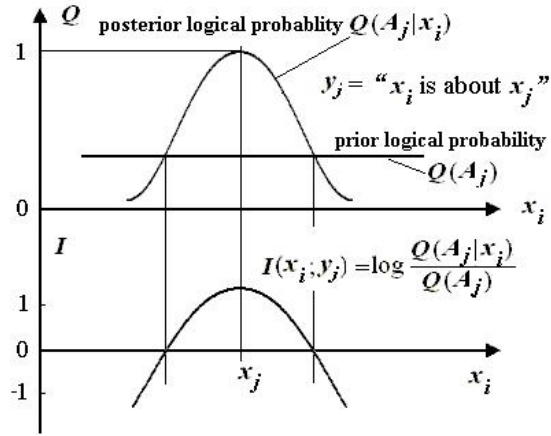

**Fig. 1.** Illustration of the generalized information formula. The more precise the prediction is, the more the information is provided. Information might be negative if the prediction is obviously wrong. The greater the error

We use an example to show the properties of the formula. Assume we need to predict a stock index for the next weekend. Let the current index be $x=100$. There are predictions $y_j=$"The index will be about $x_j$" and $y_k=$"The index will be about $x_k$". Assume there is prior knowledge:

$$Q(X) = C' \exp[-\frac{(X-x_0)^2}{2d_0^2}], \text{ where } C' \text{ is a normalizing constant};$$

$$Q(A_j | X) = \exp[-\frac{(X-x_j)^2}{2d_j^2}], \; Q(A_k | X) = \exp[-\frac{(X-x_k)^2}{2d_k^2}].$$

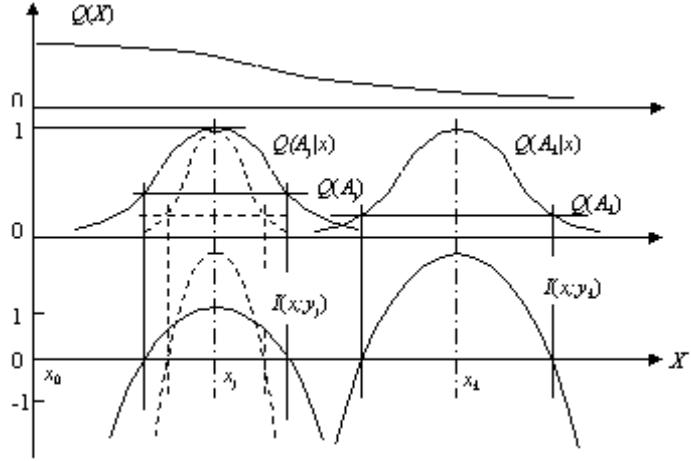

**Figure 2** Information *I* about stock index *X* conveyed by different predictions $y_j$ and $y_k$

Figure 2 shows the changes of information conveyed by $y_j$ and $y_k$ respectively with *X* changing. It tells us that the more an occasional event is correctly predicted, the more the information is. The dashed lines show the case in which $d_j$ is reduced. The corresponding prediction may be expressed as "The index will be very closed to $x_j$". It can be said that when predictions are correct, the more precise the prediction is, the more the information is. If a prediction is extremely fuzzy such as "The index will probably go up or not go up", $Q(A_j|X)$ can be represented by a horizontal line and the information will always be zero.

## 3  Comparing Information Criterion with Square Error Criterion

Assume $Q(A_j|x_i)$ is a normal function with the maximum 1, i.e.

$$Q(A_j | x_i) = \exp[-\frac{(x_i - x_j)^2}{2d^2}], \qquad (5)$$

where *d* means the precision of a prediction or the discrimination of sense organ. The less the *d* is, the higher the precision or the discrimination is. From (4) and (5), we have

$$I(x_i; y_j) = -\frac{(x_i - x_j)^2}{2d^2} - \log Q(A_j). \qquad (6)$$

If *d* and $Q(A_j)$ are 1 or constants, then the information will criterion become the square error criterion. In comparison with the square error criterion, the information criterion gives more precise predictions, or predictions that predict more occasional events, higher evaluation. If we use two criterions to evaluate people, the square error criterion means that no error is good; yet, the information criterion means that contribution over error is good.

Actually, philosopher K. R. Popper suggested using information as criterion to evaluate a scientific theory or a proposition (see page 250 in [12]) long time ago. But he didn't provide suitable information formula. The above information measure accords with Popper's theory very much [8]. If $Q(A_j|x_i)\equiv 1$, then there must be $I(x; y_i)=0$. This is just the mathematical description of Popper's affirmation that a proposition that cannot be falsified provides no information and hence is meaningless. The less a fuzzy set $A_j$ is, or the more unexpected the events in $A_j$ are, the less the $Q(A_j)$ is, and hence the bigger the $I(x; y_i)$ is while $Q(A_j|x_i)=1$. This is just the mathematical description of Popper's affirmation that a proposition with less prior logical probability has more important scientific significance if it can go though tests of facts.

For sentences "The temperature tomorrow morning will be lower than 10°C" and "There will be small to medium rain tomorrow", error is hard to be expressed because there is no center $x_j$ in fuzzy set $A_j$. However, measuring information is the same easy for given logical probability $Q(A_j|x_i)$ and probability distribution $P(x_i)$.

## 4  Generalized Kullback Formula for Sentences Selection

For given event $X = x_i$, it is easy to select a descriptive sentence $y^*$ from many sentences $y_1, y_2 \ldots$ according to the generalized information criterion. We calculate $I(x_i; y_i)$ for each $y_i$. The $y_i$ that makes $I(x_i; y_i)$ has the maximum is $y^*$ we want.

However, in general artificial intelligent systems, for given data or evidences denoted by $z$, we can only know the probability distribution $P(x_i|z)$ instead of exact event $x_i$. For example, to forecast rainfall, we first get $P(x_i|z)$ according to observed data, and then select a sentence such as "There will be heavy rain tomorrow" from many as prediction according to $P(x_i|z)$. In theses cases, we need generalized Kullback formula (see Fig. 2):

$$I(X; y_j) = \sum_i P(x_i|z) \log \frac{Q(x_i|A_j)}{P(x_i)} = \sum_i P(x_i|z) \log \frac{Q(A_j|x_i)}{Q(A_j)}, \qquad (7)$$

which is the average of $I(x_i; y_j)$ for different $x_i$. This formula is called generalized Kullback formula because it has the form of Kullback formula while $Q(x_i|A_j) = P(x_i|z)$ (for each $i$). We can prove that $I(X; y_j)$ reaches its maximum when $Q(x_i|A_j) = P(x_i|z)$.

Now, we calculate $I(X; y_j)$ for different $y_j$. The $y_j$ that makes $I(X; y_j)$ have the maximum is $y^*$ we want.

We can also use the generalized condition entropy

$$H^*(X|y_j) = \sum_i P(x_i|z) \log Q(x_i|A_j) \qquad (8)$$

as criterion to select $y_j^*$. But, actually, the calculation is not simpler than the right part of (7) because we need $Q(A_j|x_i)$ and $Q(A_j)$ to calculate $Q(x_i|A_j)$.

For language translation, we need to translate a sentence $y'$ in a language to another sentence $y^*$ in another language. In this case, we need to replace $P(x_i|z)$ with $Q(x_i|A')$, where $A'$ is a fuzzy subset of $A$, so that (7) become

$$I(X; y_j) = \sum_i Q(x_i|A') \log \frac{Q(A_j|x_i)}{Q(A_j)}. \qquad (9)$$

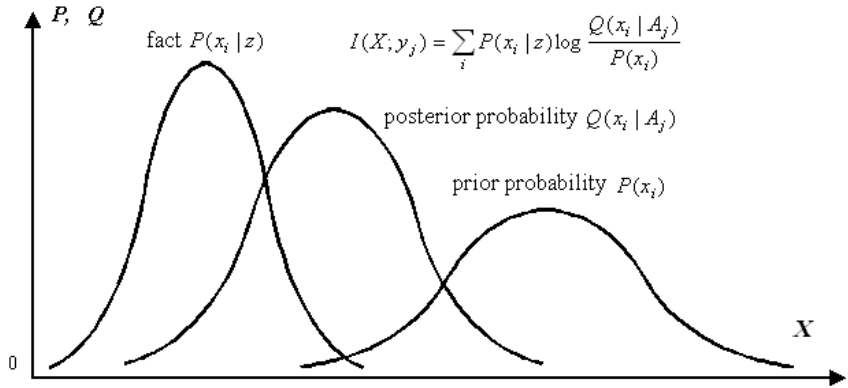

**Fig. 3.** The property of the generalized Kullback formula: the closer to the fact $P(x_i|z)$ the posterior probability $Q(x_i|A_j)$ is in comparison with the prior probability $P(x_i)$, the more the information about $X$ is conveyed by $y_j$; otherwise, the information is negative

## 4  Generalized Mutual Information Formula

Actually, the probability $P(x_i)$ in (7) may be replaced with subjectively forecasted probability $Q(x_i)$ so that we have

$$I(X; y_j) = \sum_i P(x_i|y_j) \log \frac{Q(x_i|A_j)}{Q(x_i)} \qquad (10)$$

Calculating the average of $I(X; y_j)$ for different $y_j$, we have generalized mutual information formula:

$$I(X;X) = \sum_j P(y_j) I(X; y_j) \qquad (11)$$
$$= \sum_i P(x_i, y_j) \log[Q(x_i | A_j)/Q(x_i)]$$
$$= H(X) - H(X|Y) = H(Y) - H(Y|Y),$$

where

$$H(X) = -\sum_i P(x_i) \log Q(x_j), \qquad (12)$$

$$H(X|Y) = -\sum_j \sum_i P(x_i, y_i) \log Q(x_i | A_j), \qquad (13)$$

$$H(Y) = -\sum_j P(y_j) \log Q(A_j), \qquad (14)$$

$$H(Y|X) = -\sum_j \sum_i P(x_i, y_i) \log Q(A_j | x_i). \qquad (15)$$

I call $H(X)$ forecasting entropy, which reflects the average coding length when we economically encode X according to $Q(X)$ while real source is $P(X)$, and reaches its minimum as $Q(X)= P(X)$. I call $H(X|Y)$ posterior forecasting entropy, call $H(Y)$ generalized entropy, and call $H(Y|X)$ generalized condition entropy or fuzzy entropy.

I think that the generalized information is subjective information and Shannon information is objective information. If two weather forecasters always provide opposite forecasts and one is always correct and another is always incorrect. They convey the same objective information, but the different subjective information. If $Q(X)= P(X)$ and $Q(X/A_j)= P(X/y_j)$ for each $j$, which means subjective forecasts conform to objective facts, then the subjective mutual information will be equal to objective or Shannon's mutual information.

## 5 Improving Rate Distortion Theory into Rate Fidelity Theory

Shannon proposed the rate-distortion function $R(D)$ for data compression in his creative paper [1]. For given source $P(X)$ and the upper limit $D$ of distortion

$$d(X,Y) = \sum_j \sum_i P(x_i, y_i) d(x_i, y_j), \qquad (16)$$

where $d(x_i, y_j)$ is error measure such as square error, we change channel $P(Y/X)$ to search the minimum of Shannon's mutual information $I_s(X; Y)$. The minimum denoted by $R=R(D)$ is just the rate-distortion function, which reflects necessary communication rate for given source $P(X)$ and distortion limit $D$.

Actually Shannon had mentioned fidelity criterion for lossy coding. He used the distortion, i.e. average error, as the criterion for optimizing lossy coding because the fidelity criterion is hard to be formulated. However, distortion is not a good criterion in most cases. For this reason, I replace the error function $d_{ij}=d(x_i, y_j)$ with generalized information $I_{ij}= I(x_i; y_j)$ and distortion $d(X, Y)$ with generalized mutual information $I(X; Y)$ as criterion to search the minimum of Shannon mutual information $I_s(X; Y)$ for given $P(X)=Q(X)$ and the lower limit $G$ of $I(X; Y)$. I call this criterion $I(X; Y)$ the fidelity criterion, call the minimum the rate-fidelity function $R(G)$, and call the improved theory the rate fidelity theory.

In a way similar to that in the classical information theory [14], we can obtain the expression of function $R(G)$ with parameter $s$:

$$G(s) = \sum_j \sum_i P(x_i)P(y_j)\exp(sI_{ij})\lambda_i I_{ij} \qquad (17)$$

$$R(s) = sG(s) + \sum_j \sum_i P(x_i)\log \lambda_i$$

where $s=dR/dG$ indicates the slope of function $R(G)$ (see Fig. 3) and

$$\lambda_i = 1/\sum_j P(y_j)\exp(sI_{ij}) \qquad (18)$$

In [12], I defined information value $V$ by the increment of growing speed of a portfolio because of information, and suggested to use the information value as criterion to optimize communication to get rate-value function $R(V)$, which is more meaningful in some cases.

## 6. Properties of Rate-fidelity Function and Image Compression

Now we use the information provided by different gray levels of pixels of images (see [9] for details) as sample to discuss the properties of rate-fidelity function. The conclusions are also meaningful to linguistic communication.

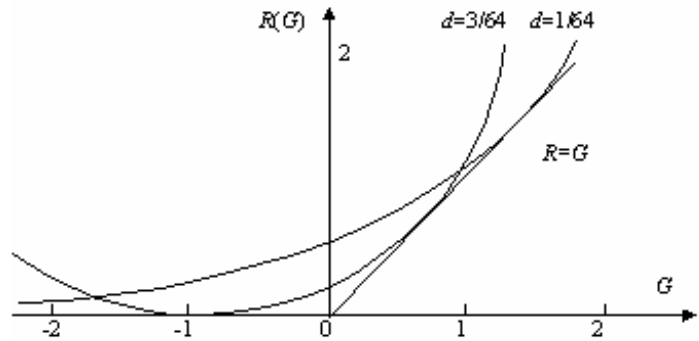

**Fig. 4** Relationship between $d$ and $R(G)$ for $b=63$

Let the gray level of a digitized pixel be a source and the gray level is $x_i=i$, $i=0, 1... b = 2^k -1$ with normal probability distribution whose expectation$=b/2$ and standard deviation$= b/8$. Assume that after decoding, the pixel also has gray level $y_j=j=0, 1... b$; the perception caused by $y_j$ is also denoted by $y_j$; and discrimination function or confusing probability function of $x_j$ is

$$Q(A_j | X) = \exp[-(X-j)^2/(2d^2)] \qquad (19)$$

where $d$ is discrimination parameter. The smaller the $d$ is, the higher the discrimination is.
Fig. 4 tells us that
1) The higher discrimination can give us more information when objective information $R$ is big enough; yet, lower discrimination is better when objective information $R$ is less. This conclusion can be supported by the fact that it is better to watch TV with less pixels or with too much snowflake-like disturbance from further distance.
  2) When $R=0$, $G<0$, which means that if a coded image has nothing to do with the original image, we still believe it reflects the original image, then the information will be negative. For linguistic communication, this means that if one believes a fortuneteller's talk, one would be more ignorant about facts and the information he has will be reduced.
  3) When $G=-2$, $R>0$, which means that certain objective information is necessary when one uses lies to deceive his enemy to some extent; or say, lies against facts are more terrible than lies according to nothing.
  4) The each line of function $R(G)$ is tangent with the line $R=G$, which means there is a matching point at which objective information is equal to subjective information, and the higher the discrimination is (the less the $d$ is), the bigger the matching information amount is. For linguistic communication, this means that for improving efficiency of communication, it is necessary to make objective information accord with subjective understanding.

5) The slope of *R*(*G*) becomes bigger and bigger with *G* increasing, which tell us that for given discrimination, it is limited to increase subjective information, and too much objective information is wasteful.

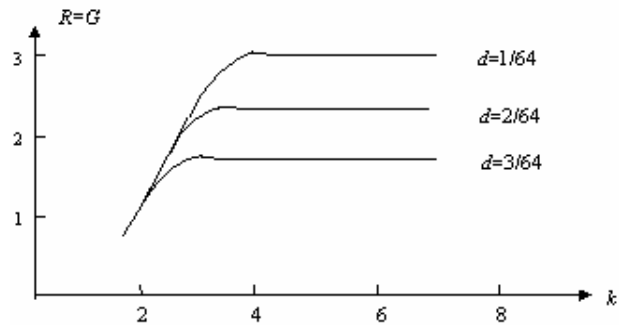

**Fig. 5.** Relationship between matching value of *R* with *G*, discrimination parameter *d*, and digitized bit *k*

Fig. 5. tells us that for given discrimination, there exists the optimal digitized bit *k'* so that the matching value of *G* and *R* reaches the maximum. If *k*<*k'*, the matching information increases with *k*; if *k*>*k'*, the matching information no longer increases with *k*. This means that too high resolution of images is unnecessary or uneconomical for given visual discrimination.

## 7  Improvement of Popper Theory

Popper and his successors tell us that reliability of a scientific proposition comes from the repeated tests by facts. What is the difference between the repeated tests and verification emphasized by logical positivism? Now we distinguish prior logical probability and posterior logical probability of a proposition. For the prior logical probability $Q(A_j)$, the less the better; yet for the posterior logical probability $Q(A_j|x_i)$, the bigger the better. So, both falsification and verification are necessary.

There are many probabilistic and fuzzy propositions, such as "High humidity will bring rain", "Thirty years old is young". How do we falsify or evaluate these propositions? Can we use a counterexample to falsify a proposition? In theses cases, the above information formula can give these propositions appropriate evaluations.

## 8 Conclusions

This paper provides the generalized information criterion, which accords to Popper's criterion of scientific advance, for sentences selection and data compression. Its rationality is supported by predictions' evaluation and many properties of rate-fidelity function.